\pdfoutput=1
\documentclass[manuscript,screen]{acmart}
\usepackage{graphicx} 
\usepackage{array} 
\AtBeginDocument{%
  }

\setcopyright{acmlicensed}
\copyrightyear{2025}
\acmYear{2025}
\acmDOI{XXXXXXX.XXXXXXX}
\acmConference[CHI Play '25]{The Annual Symposium on Computer-Human Interaction in Play }{October 13--16,
  2025}{Pittsburgh, PA}
\acmISBN{978-1-4503-XXXX-X/2025/10}

\begin{document}

\title{Integrating Artificial Intelligence as Assistive Technology for Older Adult Gamers: A Pilot Study }

\author{Yichi Zhang}
\email{zhang.yichi6@northeastern.edu}
\author{Brandon Lyman}
\email{lyman.br@northeastern.edu}
\author{Celia Pearce}
\email{c.pearce@northeastern.edu}
\author{Miso Kim}
\email{<m.kim@northeastern.edu>}
\author{Casper Harteveld}
\email{c.harteveld@northeastern.edu}
\author{Leanne Chukoskie}
\email{l.chukoskie@northeastern.edu}
\author{Bob De Schutter}
\email{b.deschutter@northeastern.edu}
\affiliation{%
  \institution{College of Arts, Media and Design, Northeastern University}
  \city{Boston}
  \state{Massachusetts}
  \country{USA}
}

\renewcommand{\shortauthors}{Zhang et al.}

\begin{abstract}
With respect to digital games, older adults are a demographic that is often underserved due to an industry-wide focus on younger audiences’ preferences and skill sets. Meanwhile, as artificial intelligence (AI) continues to expand into everyday technologies, its assistive capabilities have been recognized, suggesting its potential in improving the gaming experience for older gamers. To study this potential, we iteratively developed a pilot survey aimed at understanding older adult gamers' current gameplay preference, challenges they are facing, and their perspectives of AI usage in gaming. This article contributes an overview of our iterative survey-design workflow, and pilot results from 39 participants. During each iteration, we analyzed the survey’s efficacy and adjusted the content, language,  and format to better capture meaningful data, and was able to create a refined survey for a larger, more representative future parent study. At the same time, preliminary findings suggest that for older adult gamers, usability issues in gaming remain key obstacles, while this demographic's perceptions of AI are shaped by both its practical benefits and concerns about autonomy and complexity. These findings also offer early insights for the design of age-inclusive, AI-supported gaming experiences.
\end{abstract}



\received{4 June 2025}
\received[revised]{d mmm yyyy}
\received[accepted]{d mmm yyyy}

\maketitle

\section{Introduction}
Recent surveys have shown that older adults now represent a growing demographic in digital gaming \cite{KakullaGamesShould2024, KakullaGamers50Plus2023}. Despite this growth, most games are still designed with younger players in mind \cite{PetersonAgeismissue2018} or puts older adult gamers in the position of "gaming to fight aging" \cite{WuOlderGamer2020a}. As a result, mismatches appear between older adults’ needs and what games offer, introducing various challenges to older players while navigating some games that are not tailored to their abilities or preferences. 
AI has become a powerful tool across various domains, offering older adults with support through personalized adaptation, real-time feedback, daily assistance and so on. \cite{NaseerIntegratinggenerative2025,VrancicRoleSmart2024}. Yet, in the world of digital games, an increasingly important site of entertainment, learning, and social interaction for older adults, the use of AI remains underexplored. This raises an important question: \textit{how might AI be designed to support older adults in digital games—not by replacing player interaction, but by lowering barriers to play?}
To begin tackling this greater question, we first need to understand deeply this demographic’s current gaming experience and their willingness to use AI as a potential solution to enhance gameplay. However, capturing these insights requires careful methodological consideration. Older adult gamers are a group with diverse interests, thus designing an instrument that neither leads responses nor makes assumptions about their abilities is a challenge.
This paper details a pilot for a parent study with the research questions as:

\begin{enumerate}
    \item \textit{What are the gaming preferences of older gamers (i.e., what types of games do they want to play)?}
    \item \textit{What challenges do older gamers face when playing digital games?}
    \item \textit{How do older gamers perceive AI in the context of digital games?}
\end{enumerate}

The importance of pilot studies in the intersection of HCI and digital games has been evidenced by previous literature \cite{MohaddesiImportancePilot2020}. To ensure meaningful and reliable data from older adult gamers, this pilot study also addresses an implicit but essential research question: 
\textit{how can we design an effective questionnaire to capture older adults gamers' experience and thoughts? }

The purpose of this pilot is manifold: (1) testing the practicality of our methods for the parent study \cite{CopeConductingpilot2015}; (2) tailoring the contents of our instruments to the needs and tendencies of our target population \cite{WildAreyou2016, LenznerCognitiveburden2010, FarageDesignPrinciples2012, WeirOlderAdults2025} and (3) providing evidence that the selected methods will generate results that help to answer the research questions. This pilot study contributes to an adaptive and iterative approach to survey design for older adult gamers. At the same time, it offers preliminary insights related to the broader research goals: it identifies a catalog of games currently preferred by older adult gamers, affirms the applicability of known gameplay challenges to this demographic, and sheds light on their general attitudes toward AI technologies in gaming.

\section{Background}
\textit{Gaming Preferences of Older Adults.} Nowadays, gaming has become a lifelong activity \cite{DeSchutterNeverToo2011a}. Many individuals who played games when they were younger continue to do so as they age, contributing to the expansion of older adult players as an important demographic in the gaming industry \cite{BunzExaminingyounger2020a}. 
Research on older adult gamers shows a wide range of preferences. While some enjoy fast-paced, competitive games, most prefer slower, cognitively focused genres such as puzzle, card, and strategy games \cite{CheshamWhatOlder2017a, BlockerGamingpreferences2014a, SecerDigitalGaming2023}. Key motivations include maintaining mental sharpness \cite{DeSchutterOlderPlayer2014a, CardonaMeaningfullearning2023}, learning new skills \cite{EsnardOlderpeople2024}, fostering social connections with friends and family \cite{KaufmanOlderAdults2016d, CardonaMeaningfullearning2023}, seeking for relaxation and escapism \cite{DeSchutterOlderPlayer2014a, PearceTruthBaby2008}, among others. As older adults increasingly turn to games for cognitive, social, and emotional enrichment, identifying their gaming challenges for the purpose of mitigation becomes a crucial foundation to promote inclusive and enjoyable gaming experiences.

\textit{Gaming Challenges of Older Adults.} During the natural aging process, an individual's physical and cognitive functions may decline to varying degrees \cite{MurmanImpactAge2015,CooperAgeGender2011}. Research has revealed a series of recurring gaming challenges stemming from age-related changes, age-unfriendly game usability designs, and common communication issues in digital space. Table \ref{tab:challenges} shows the major challenges widely recognized in the literature, categorized by the nature of the challenge, with a mnemonic code and short description for ease of reference.

\begingroup
\def\arraystretch{1.5}
\begin{table*}
    \tiny
    \centering
    \begin{tabular}{ |m{1cm}|m{3cm}|m{5.5cm}|m{2.5cm}|m{1cm}|} 
        \hline
        \textbf{Code} & \textbf{Category} & \textbf{Challenge Description} & \textbf{Short Description} & \textbf{Source} \\
        \hline
        \textbf{CH1} & Sensory and Physical Challenges & Hard to read text or unclear fonts & Text Legibility & \cite{GerlingGameDesign2012a, GamberiniCognitiontechnology2006, JohnsonAppendixDesign2017a}\\
        \textbf{CH2} & Sensory and Physical Challenges & Difficulty seeing colors or icons & Visual Clarity & \cite{IjsselsteijnDigitalgame2007a, GerlingGameDesign2012a, GamberiniCognitiontechnology2006}\\
        \textbf{CH3} & Sensory and Physical Challenges & Difficulty using controllers, keyboards, or touch screens  & Input Challenges & \cite{IjsselsteijnDigitalgame2007a,GerlingGameDesign2012a,JohnsonAppendixDesign2017a, VasconcelosDesigningtabletbased2012a}\\
        \textbf{CH4} & Usability Design Issues & Confusing or hard-to-use game menus or navigation  & Menu Usability & \cite{IjsselsteijnDigitalgame2007a, JohnsonAppendixDesign2017a, LeeMobileGame2021a}\\
        \textbf{CH5} & Usability Design Issues & The game fails to provide enough in-game assistance or tutorials, or the tutorials are unhelpful or overwhelming & Tutorial Quality  & \cite{BootOlderadults2020d, LeeMobileGame2021a}\\
        \textbf{CH6} & Cognitive Challenges & Difficulty remembering control and in-game mechanics & Memory Load & \cite{IjsselsteijnDigitalgame2007a, GerlingGameDesign2012a, CotaMobilegame2015}\\
        \textbf{CH7} & Cognitive Challenges & Difficulty making quick decisions under time pressure & Time Pressure & \cite{GamberiniCognitiontechnology2006, GuptaDesigningSerious2013}\\
        \textbf{CH8} & Communication Issues & Difficulty communicating in multiplayer games & Multiplayer Communication & \cite{NapSeniorgamers2009a, JanssenOlderAdults2023}\\
        \textbf{CH9} & Communication Issues & Online toxicity, or negative online social experiences  & Online Toxicity & \cite{MattinenOnlineAbuse2018,ZsilaToxicbehaviors2022}\\
        \hline
    \end{tabular}
    \caption{Challenges faced by older adults in digital games, organized by category}
    \label{tab:challenges}
\end{table*}
\endgroup

In response to the challenges, the field of game accessibility has made notable progress in recent years. In addition to research-informed guidelines, features such as adjustable text/icons, and sound levels, color-blind modes, etc. have become common components of industry accessibility standards \cite{GameAccessibility,SIGguidelines}. These features not only broaden access for players with disabilities but also help address some of the sensory-related barriers faced by older gamers, helping to mitigate the risk of further cognitive decline associated with sensory impairments \cite{NagarajanVisionimpairment2022}. 
However, while sensory support can often be achievable through static accessibility features, cognitive challenges such as managing fast-paced gameplay or high information load can be more difficult to accommodate. These issues may require more dynamic and context-sensitive support, which adaptive AI technologies are positioned to provide.

\textit{Older Adults' Perceptions of AI.} While AI is increasingly integrated into everyday life, a Foresight 50+ survey showed that only 17\% of adults aged 50 and over are very familiar with AI \cite{OlderAdults2023}. Older adults are beginning to recognize the benefits of AI tools in areas such as healthcare and home assistance, but concerns remain around accessibility, privacy, autonomy, and trust \cite{ShiwaniNewHorizons2023, PadhanArtificialIntelligence2023, ShandilyaUnderstandingOlder2024}. Regarding AI in gaming, dedicated efforts to understand older adult gamers’ attitudes as a specific demographic remain limited. 
Some studies on general age groups have shown that players’ attitudes toward AI in gaming are relatively neutral, depending on the role of AI in the game. When AI acts as a tool and provides customization, meaningful interactions, and in-game assistance, it can enhance players' satisfaction; when AI plays the role of competitor or decision maker, players' reactions are more complicated: some appreciate the challenge and realism, while others worry about the fairness and unpredictability of AI \cite{ButtLetsplay2021, KarakusExaminingAttitudes2023, SparrowEthicalAI2024}. These general findings may also be relevant for understanding older adult gamers. Given that perceived practicality and ease of use significantly influence older adults’ openness to AI \cite{ShadeEvaluatingOlder2025, ShandilyaUnderstandingOlder2024}, AI may be more acceptable to them when performing as an assistive tool in games. Besides, compared with younger users, older adults tend to have lower trust in the reliability of AI and express more concerns about losing control over decision-making \cite{WongExploringOlder2025,ShiwaniNewHorizons2023}, which also highlighting the importance of ensuring transparency and supporting autonomy when designing AI-enhanced game systems.

\textit{Instruments.} We considered several instruments and techniques from literature that could help to answer our research questions. Pearce used a 23 item, 5-point Likert-scale genre matrix to collect information on genre preferences of Baby Boomers (born between 1943 and 1968) \cite{PearceTruthBaby2008}. The Intrinsic Motivations to Gameplay (IMG)  is a validated, 15-item inventory that can be used to collect data on motivational rationale for playing digital games, which could indicate player preferences \cite{VahloFiveFactorInventory2019a}. 

\section{Methods}
To better understand the gaming preferences, the gaming challenges, and the perception of AI within the older adult gamer demographic, we piloted an in-depth, online questionnaire to survey older adult gamers and begin answering our research questions. 

\subsection{Study Participants}
Based on previous research, we concluded that 30 participants is an appropriate, minimum sample size for this pilot study \cite{PernegerSamplesize2015}. Our inclusion criteria were that the participants needed to be aged 50+ and have been playing video games at least once a week for the past 6 months. We consider people who meet these criteria to be “older adult gamers.”
Participants were recruited through the Prolific platform and were paid \$10.00 USD for completing the survey. A total of 39 older adult gamers responded to our survey across four iterations. 
Ages of our participants ranged from 50-68 ($\mu$ = 55.07, sd = 4.89). The proportion of females in the sample was 56.41\% (F=22, M=17). White / Caucasian people are overrepresented in the sample (79.49\%). Based on these demographic results, we conclude that the parent study will need to move beyond Prolific in order to establish a representative sample. Our study protocol was reviewed and approved by our university’s internal review board.

\subsection{Survey Structure and Iteration}
We adopted an iterative process to refine our questionnaire for use in the parent study \cite{Nielsenmathematicalmodel1993}. The iterations detailed in this section resulted in a questionnaire consisting of six sections: (1) consisted of consent and screener questions, (2) collected digital gameplay habits of older adults (to be explored in the parent study), (3) collected information on favorite games, (4) focused on perceptions of AI and included the General Attitudes towards Artificial Intelligence Scale (GAAIS) \cite{SchepmanInitialvalidation2020a} (to be explored in the parent study), (5) captured challenges older adult gamers face when playing games, and (6) collected demographic information about the participants.

The iterations were used to assess five key areas: (1) the usability of the questionnaire design, (2) the clarity of the questions asked, (3) accuracy, (4) depth of responses, and (5) duration, aiming to minimize survey length. The iterative setup allowed us to respond to issues in these five key areas without wasting resources; for example, paying participants to answer questions that ultimately get deleted. After each iteration, we reviewed the responses and adjusted the questionnaire to improve it in one of those five key areas. Here we cover some example adjustments made regarding each.

\textbf{Usability.} We noticed the over-reliance on open-ended questions may have been fatiguing for our participants, as indicated by short or non-existent responses, so we focused on collecting common answers and converting open-ended questions into multiple choice questions.

\textbf{Accuracy.} We initially included a question about non-game related hobbies, believing that such a question could lead to insight about game preferences. Based on the responses, it did not seem like we were getting such insight, and we eliminated the question as it proved irrelevant to answering our research questions.

\textbf{Clarity.} Our participants did not seem to know what we meant when we asked about “assistive technologies” as an open-ended question. To remedy this, we converted the question into a multiple choice and listed many examples of what we mean by assistive technologies (i.e., screen readers, adaptive controllers). We also gave participants the option to write in an assistive technology not listed. This elegantly encouraged our participants to provide insights on the types of technologies we were interested in, improving the clarity of the questionnaire.

\textbf{Depth of Response.} We noticed that many of our open-ended questions were able to be answered with a “yes” or a “no.” Even though we prompted the participant for more insight as to why, most participants stopped after their single-word response. We adjusted the open-ended questions to avoid single-word responses altogether and encourage deeper reflection.

\textbf{Duration.} Many of the above changes also reduced the duration of the questionnaire, as multiple-choice questions are generally quicker to answer than open-ended questions. Initially we considered and implemented the Intrinsic Motivations to Gameplay (IMG) \cite{VahloFiveFactorInventory2019a} to capture motivation data of our participants. As a method of reducing duration, we narrowed our scope to omit motivation as it has been covered extensively in the literature. As a result, IMG was no longer needed, and removing the 15-item Likert-scale matrix was likely to have an impact on the duration. The average time taken for the final iteration of the survey was about 28 minutes.

A brief summary of changes between iterations can be found in Table \ref{tab:iterations}.

\begingroup
\def\arraystretch{1.5}
\begin{table*}
    \tiny
    \centering
    \begin{tabular}{ |m{1.25cm}|m{3.9cm}|m{3.9cm}|m{3.9cm}|} 
        \hline
        \textbf{Version} & \textbf{Key Features} & \textbf{Additions/Changes from Last Iteration} & \textbf{Limitations} \\
        \hline
        \textbf{Initial Survey} & Used validated instruments (IMG, GAAIS); Used adapted version of Pearce’s 2008 genre matrix; mixed question types (multiple choice, open-ended); included questions on AI attitudes. & n/a & Did not capture gameplay challenges clearly; too many open-ended questions; genre matrix was cumbersome; assistive tech and favorite games questions were unclear. \\
        \textbf{Iteration 1} & Structured favorite games section; reduced number of open-ended questions. & Collected 3–6 favorite games with reasons; converted repetitive open-ended responses into multiple-choice; clarified assistive technology question with examples. & Some challenge and AI questions remained confusing; some questions were overly specific; participants often responded with "yes" or "no" to open-ended prompts. \\
        \textbf{Iteration 2} & Focused on gameplay challenges; added two Likert scales for frequency and extent of challenges. & Assessed frequency and extent of challenges; clarified AI-related questions; incorporated participant write-ins into multiple-choice options. & Survey still too long (average 32 min); limited insight from hobby questions; gameplay motivation deemed out of scope. \\
        \textbf{Iteration 3} & Streamlined content and reduced length. & Removed IMG and hobby questions; added participant-suggested challenge item; finalized clarity and grammar edits. & Final version completed \\
        \hline
    \end{tabular}
    \caption{Summary of survey iterations}
    \label{tab:iterations}
\end{table*}
\endgroup

\section{Results}
\subsection{Older Adult Gamers: Gaming Preferences}
We required participants to provide at least 3 of their favorite games, and gave participants the option to provide up to three more should they desire to do so. 32.14\% of participants chose to provide more than three games. Since we asked participants to indicate their favorite games in all four versions of the questionnaire, we derived valuable insights by organizing and analyzing the compiled responses. As a result, we generated a list of 148 games total, with 113 unique entries among the 39 participants.
For preliminary analysis, the games were categorized into broad genres similar to those presented in Pearce’s study \cite{PearceTruthBaby2008}. We then calculated the percentage of participants that reported each genre at least once in their list of favorite games. The top five genres are presented in Table \ref{tab:genres}.

\begingroup
\def\arraystretch{1.5}
\begin{table*}
    \tiny
    \centering
    \begin{tabular}{ |m{4.3cm}|m{4.3cm}|m{4.3cm}|} 
        \hline
        \textbf{Genre} & \textbf{\% of Participants Reporting a Favorite Game in the Genre (Count)} & \textbf{Most Popular Example (Count)}\\
        \hline
        RPG/Adventure & 33.33\% (13) & The Legend of Zelda: Breath of the Wild (3) \\
        Puzzle & 33.33\% (13) & Candy Crush Saga (8) \\
        Shooter & 23.07\% (9) & Call of Duty Series (6) \\
        Sports Sim & 20.51\% (8) & Various\textsuperscript{†} (1) \\
        Strategy & 20.51\% (8) & Civilization Series (3) \\
        \hline
    \end{tabular}
    \caption{Top 5 Genres based on \% of participants reporting at least one favorite game in the genre. †Note: Each example of the Sports Sim genre was reported once (Dream League Soccer, FIFA Series, Golf It!, Golf Clash, Madden Series, NBA 2K Series, NFL 2K Series, NHL Series, Pro Feel Golf)}
    \label{tab:genres}
\end{table*}
\endgroup

Other findings of note are that 82.05\% of participants used smartphones or tablets to play digital games,underscoring the importance of \textit{ensuring the online questionnaire is optimized for mobile phone use}. Ensuring multiple choice questions are displayed vertically instead of horizontally, minimizing Likert-scale matrix questions, and cutting short-answer questions (or making them optional) can help older adult gamers comfortably respond to the questionnaire. Interestingly, only 15.38\% of participants reported at least one MMORPG as a favorite game.
In addition to asking the participants to indicate their favorite games, we asked participants for a reason the provided game was their favorite in the form of an open-ended question. The key observation here is that many participants responded with in-depth justification for their favorite games. For example, P15 provided descriptive writeups spanning many elements of gameplay, including social connections (“I still play games regularly with friends I met playing this game 20+ years ago”). Some responses indicated personal or emotional connections with the game, such as P7 on World of Warcraft, “It was the very first game that I started playing after my husband died, it helps me relax and regroup.” Others indicated that playing their favorite games connects them to loved ones, such as P36 on Lego Star Wars, “I can easly [sic] play the game with my grandchildren. It encourages teamwork, helping you solve puzzels [sic] and defeat enemies together.” We learned from this question in the survey iteration that \textit{this demographic loves to talk about their favorite games and generally will respond deeply to open-ended questions about their favorite games}. Further evidence of this is the proportion of participants who chose to give more than the required three games (32.14\%).

\subsection{Older Adult Gamers: Gaming Challenges}
We consolidated the findings from the research and presented common challenges to participants in the questionnaire, asking them how often they experienced the challenges (5-point Likert-scale; 1 - Never, 5-Always) and to what extent the challenges negatively impacted their gameplay experience (7-point Likert-scale; 1-Not at all, 7- Extremely). Although we acknowledge the switch from a 5 to 7-point Likert-scale was unintentional, but the extra granularity in extent allows us to capture more nuance in a more abstract metric. In addition, we used an open-ended question to allow participants to report on challenges we did not capture in the questionnaire. Refer to Table \ref{tab:challenges} for each identified challenge and its source.

We began asking participants about specific challenges in the second iteration of the questionnaire; therefore, we gathered 20 responses on this series of questions. Table \ref{tab:freqext} shows the average frequency of the 9 challenges alongside the average extent to which those challenges were indicated to negatively impact gameplay experience.

\begingroup
\def\arraystretch{1.5}
\begin{table*}
    \tiny
    \centering
    \begin{tabular}{ |m{3.6cm}|m{4.4cm}|m{2.5cm}|m{2.5cm}|} 
        \hline
        \textbf{Code} & \textbf{Category} & \textbf{Frequency ($\mu$,sd)} & \textbf{Extent ($\mu$,sd)}\\
        \hline
        \textbf{CH1} (Text Legibility)  & Sensory and Physical Challenges & 2.00, 1.41 & 2.06, 1.43 \\
        \textbf{CH2} (Visual Clarity) & Sensory and Physical Challenges &1.40, 1.27 & 1.40, 1.18 \\
        \textbf{CH3} (Input Challenges) & Sensory and Physical Challenges &0.95, 1.15 & 1.69, 1.44 \\
        \textbf{CH4} (Menu Usability)  & Usability Design Issues &1.90, 1.41 & 2.44, 1.46 \\
        \textbf{CH5} (Tutorial Quality)  & Usability Design Issues &\textbf{2.05}, 1.36 & \textbf{2.61}, 1.79 \\
        \textbf{CH6} (Memory Load)  & Cognitive Challenges &1.90, 1.41 & 2.18, 1.55 \\
        \textbf{CH7} (Time Pressure)  & Cognitive Challenges &1.80, 1.28 & 1.67, 1.19 \\
        \textbf{CH8} (Multiplayer Communication) & Communication Issues &1.30, 1.34 & 1.62, 1.19 \\
        \textbf{CH9} (Online Toxicity) & Communication Issues &2.00, 1.75 & 2.36, 1.65 \\
        \hline
    \end{tabular}
    \caption{Frequency and Extent of Gameplay Challenges}
    \label{tab:freqext}
\end{table*}
\endgroup

The challenges reported to be most frequently encountered were tutorial quality (CH5, $\mu$ = 2.05, sd = 1.36), text legibility (CH1, $\mu$ = 2, sd = 1.41), and online toxicity (CH9, $\mu$ = 2, sd = 1.75). Regarding categories, usability design and cognitive related challenges were scored with relatively high frequencies. The challenges that negatively impacted gameplay experience the most were reported as tutorial\textbf{ }quality (CH5, $\mu$ = 2.61, sd = 1.79), menu usability (CH3, $\mu$ = 2.43, sd = 1.46), and online toxicity (CH9, $\mu$ = 2.35, sd = 1.65). Again, challenges brought by age-unfriendly usability designs were rated as hindering game play the most. Note, a 10th common challenge was identified through a write-in response from P24, which read “Gets too hard or challenging that I cannot pass.”  From this we identified the \textbf{difficulty curve} challenge, which we will collect more data on in the parent study.

\subsection{Older Adult Gamers: Opinions of AI}
\begin{figure*}[ht]
\centering
\includegraphics[width=0.75 \textwidth]{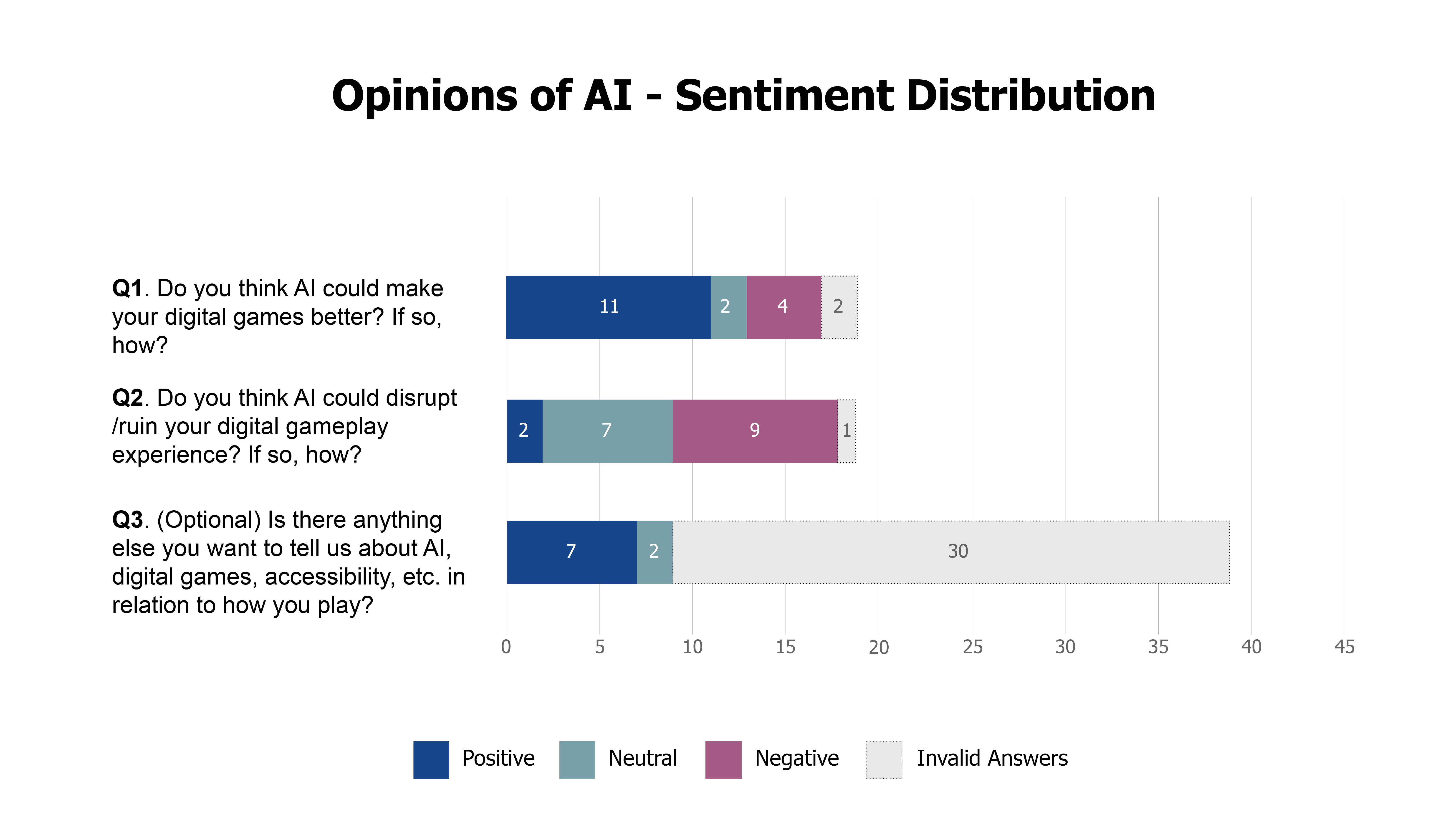}
\caption{Sentiment distribution from open-ended questions on opinions of AI}
\Description[Test]{Test}
\label{fig:aiopinions}
\end{figure*}

We retained three open-ended questions capturing older adults’ attitude towards AI, in the gaming contexts. We realized during the iteration that \textit{reminding participants about what AI can do by recounting existing AI technologies helped them to make sense of these more abstract open-ended questions in the questionnaire}. After excluding responses that lacked clear attitudes or substantive insights (e.g., “I’m not sure,” “I can’t think of anything”), the remaining valid responses were categorized as positive, neutral, or negative based on the underlying sentiment, regardless of whether the original question was framed positively or negatively. An overview of the distribution of responses across sentiment categories is provided in Figure \ref{fig:aiopinions}.

Participants were generally optimistic about AI’s potential to improve digital games, often describing scenarios from their own experiences and preferences. P16 wrote, "I think AI in a game like Skyrim would be actually amazing. Interacting with NPCs attached to a real AI would totally change how we view games." Among responses, frequently mentioned potential of AI included enhancing NPC behavior (n = 5), improved game graphics (n = 2), providing customizable settings or generating adaptive opponent AI through player behavior learning (n = 3) and offering strategic advice or performing repetitive tasks (n = 2).
 
However, participants who expressed concern about AI in games pointed to potential risks. Among these cautious responses, the common theme emerged as poor content quality, including creating boring or generic NPCs, (n = 3) and overly difficult AI opponents due to behavior learning (n = 3), as P17 answered, "Maybe, because they probably would always win." Other concerns include technical issues like algorithmic or coding errors (n = 4), and unwanted intrusions like ads, tracking, or malware (n = 3).
 
For the optional open-ended question, valid responses again showed positive outlooks of AI in improving gaming. For example, P16 wrote, "I definitely think AI is going to do some amazing things. for accessibility alone AI will be huge. Imagine if it can give verbal clues to the player as to what is going on on the screen? Or more tactile controls that are controlled by AI that translates what is happening to the device". In summary, responses highlighted the potential for AI to enhance accessibility (n = 2), support social connection (n = 1), and improve tutorials (n = 1). Still, across all questions, a few participants voiced a general dislike of AI technology a total of 3 times, regardless of context.

\section{Discussion}
\textit{Gaming Preferences.} As we iterated upon the questionnaire and implemented what we learned, we generated a set of preliminary results to further motivate the parent study.
Our findings differ from those of Pearce’s 2008 study. In her study, she reported that about 66\% of participants listed roleplay/mystery/adventure as their preferred genre \cite{PearceTruthBaby2008}. Our analysis showed a slight decrease in interest in this genre, with only about 36\% of participants reporting RPG/Adventure games or Mystery games as one of their favorite games. Only 23\% of our participants reported a shooter to be one of their favorite games, which is far less than Pearce’s figure of 60\%. Such discrepancies may exist because (1) we did not use the exact same method as Pearce, as we asked for favorite games instead of genres alone, and (2) people in this age group were different at the time of Pearce’s study. At present, it is unclear if this difference is related to age or the evolution of games in the past 17 years.

\textit{Gaming Challenges.} While participants reported varying gaming habits and preferences, the most common challenge was tutorial quality (CH5) and text legibility (CH1), adhering to findings in prior literature. Beyond the relatively recent attention to tutorial design as a specific concern for older gamers \cite{BootOlderadults2020d, LeeMobileGame2021a}, frequently reported issues relating to sensory decline and its impact on gameplay were reaffirmed \cite{GerlingGameDesign2012a, GamberiniCognitiontechnology2006, JohnsonAppendixDesign2017a, IjsselsteijnDigitalgame2007a}, highlighting the fundamental importance of sensory support mechanisms. In terms of the challenge extent, tutorial quality (CH5) and menu usability (CH4) stood out, emphasizing that non-inclusive usability design remains a significant barrier to older adults' digital gaming experiences. While accessible interfaces and navigation designs for older adults have been the focus of ongoing improvement over the past two decades \cite{IjsselsteijnDigitalgame2007a, JohnsonAppendixDesign2017a, LeeMobileGame2021a}, tutorial design remains a more complex issue. Game tutorial design and efficacy often vary greatly depending on genre and individual player, making it difficult to develop a universal solution. In this regard, besides holding great potential for enabling personalized user interfaces, AI can also provide unique possibilities of tutorial support, by tailoring instructional content to a variety of game genres, players’ skill sets and pace of learning.

\textit{Opinions of AI in Games.} Attitudes toward the use of AI in games were nearly evenly distributed among older adult gamers. Participants expressed clear preferences regarding the roles AI should play. Consistent with prior findings \cite{ButtLetsplay2021, KarakusExaminingAttitudes2023, SparrowEthicalAI2024}, they tended to prefer AI as a supportive tool that enhances gameplay and assists players, rather than one that introduces additional challenges or complexity. Concerns about AI interfering with human control or autonomy were also raised repeatedly, reaffirming earlier literature that identified these issues as older adults’ hesitation with AI usage, even within entertainment contexts \cite{WongExploringOlder2025, ShiwaniNewHorizons2023}. In summary, these results suggest that the integration of AI into games for older adults must prioritize practical utility, transparency, and the preservation of user autonomy.

\section{Limitations and Future Work}
This pilot study has several limitations. First, it did not include questions related to auditory challenges, which may be relevant for older adult gamers’ sensory changes. Second, the iterative nature of survey development led to inconsistent response counts across items, limiting cross-item comparisons. Third, the sample was recruited primarily via Prolific, resulting in limited demographic diversity. Most participants identified themselves as White and reported higher income levels, which constrains data generalizability.

Building on this pilot study, the parent study will use a finalized, unified version of the survey to improve data consistency and enable more robust analyses. It will also expand recruitment through lifelong learning institutes, online communities, and social media to reach a more diverse older gamer population. These refinements aim to provide deeper insight into older gamers’ experiences, expectations and concerns of AI regarding gaming contexts, and to gather understandings on how AI can support more accessible, engaging, and age-inclusive digital gameplay.

\bibliographystyle{unsrt}
\bibliography{References}
\end{document}